\documentclass{article}

\usepackage{arxiv}

\usepackage[utf8]{inputenc} 
\usepackage[T1]{fontenc}    
\usepackage{hyperref}       
\usepackage{url}            
\usepackage{booktabs}       
\usepackage{amsfonts}       
\usepackage{nicefrac}       
\usepackage{microtype}      
\usepackage{lipsum}
\usepackage{graphicx}

\title{WI Fast Stats: a collection of web apps for the visualization and analysis
  of WI Fast Plants data}

\author{
  Yizhou Liu\\
  Department of Computer Science\\
  University of Wisconsin-Madison\\
   \And
  Claudia Sol\'is-Lemus\thanks{Corresponding author: solislemus@wisc.edu}\\
  Wisconsin Institute for Discovery\\
  and Department of Plant Pathology\\
  University of Wisconsin-Madison\\
}

\begin{document}
\maketitle

\begin{abstract}
  WI Fast Stats is the first and only dedicated tool tailored to the
  WI Fast Plants educational objectives.  WI Fast Stats is an
  integrated animated web page with a collection of R-developed web
  apps that provide Data Visualization and Data Analysis tools for WI
  Fast Plants data.  WI Fast Stats is a user-friendly easy-to-use
  interface that will render Data Science accessible to K-16 teachers
  and students currently using WI Fast Plants lesson plans.  Users do
  not need to have strong programming or mathematical background to
  use WI Fast Stats as the web apps are simple to use, well
  documented, and freely available.
\end{abstract}


\section{Summary}

WI Fast Plants (\url{https://fastplants.org/}) \cite{williams1986rapid} were developed as a research tool at the University of Wisconsin-Madison and have been used by K-16 teachers around the world for nearly 30 years as an educational model-organism. As the name suggest, WI Fast Plants have a short time from planting to flowering (about 2 weeks), and thus, these plants are used in classrooms to engage learners of all ages and grade levels into the investigation of plant life cycles, the growth of flowers, the role of environmental factors on plants, the energy transformation of plants, the genetics of hybrids among many other topics.

Thousands of students from elementary school to college level grow WI Fast Plants in class or at home, and collect data from their experiments, yet they do not have any user-friendly tool to visualize or analyze these data. Existing Data Science platforms such as CODAP \cite{codap} are either too complicated to use or expensive. Furthermore, existing interfaces are not tailored to WI Fast Plants data or educational objectives, and lack the flexibility to evolve alongside the lesson options provided by WI Fast Plants.

\section{Statement of need}

WI Fast Stats (\url{https://wi-fast-stats.wid.wisc.edu/}) is an integrated animated web page which serves as a medium to a collection of R-developed web apps that provide Data Visualization and Data Analysis tools for WI Fast Plants data. 
Each web app corresponds to a educational unit linked to a specific WI Fast Plants webinar and it serves two main functions: 1) K-16 teachers attending the WI Fast Plants webinar will learn how to design Data Science exercises through the web app for their students and 2) students can use the web app independently to learn about visualization and analysis via the publicly available sample datasets and educational materials.

In this sense, WI Fast Stats is designed to be used by both K-16 teachers and by
students from elementary school to college. A preliminary version of a web app has already been used in a 2020 WI Fast Plants webinar with over 300 high-school teachers across the U.S. in attendance. Based on the comments at the webinar and follow-up messages, the attendees were deeply impressed by the potential of the web app to revolutionize the manner in which they currently teach Data Science in the classroom. 

WI Fast Stats is a user-friendly easy-to-use interface that will render Data Science accessible to teachers and students without strong programming or mathematical background. Because WI Fast Stats was also created and is maintained at the University of Wisconsin-Madison, it is the perfect companion to the already widely successful educational tool of WI Fast Plants. WI Fast Stats will evolve alongside WI Fast Plants by fulfilling the Data Science needs of the thousands of teachers and students around the world who currently use WI Fast Plants to learn the complexities of plants and the biological world.

\section{Description of the website}

\begin{figure}
    \centering
    \includegraphics[scale=0.25]{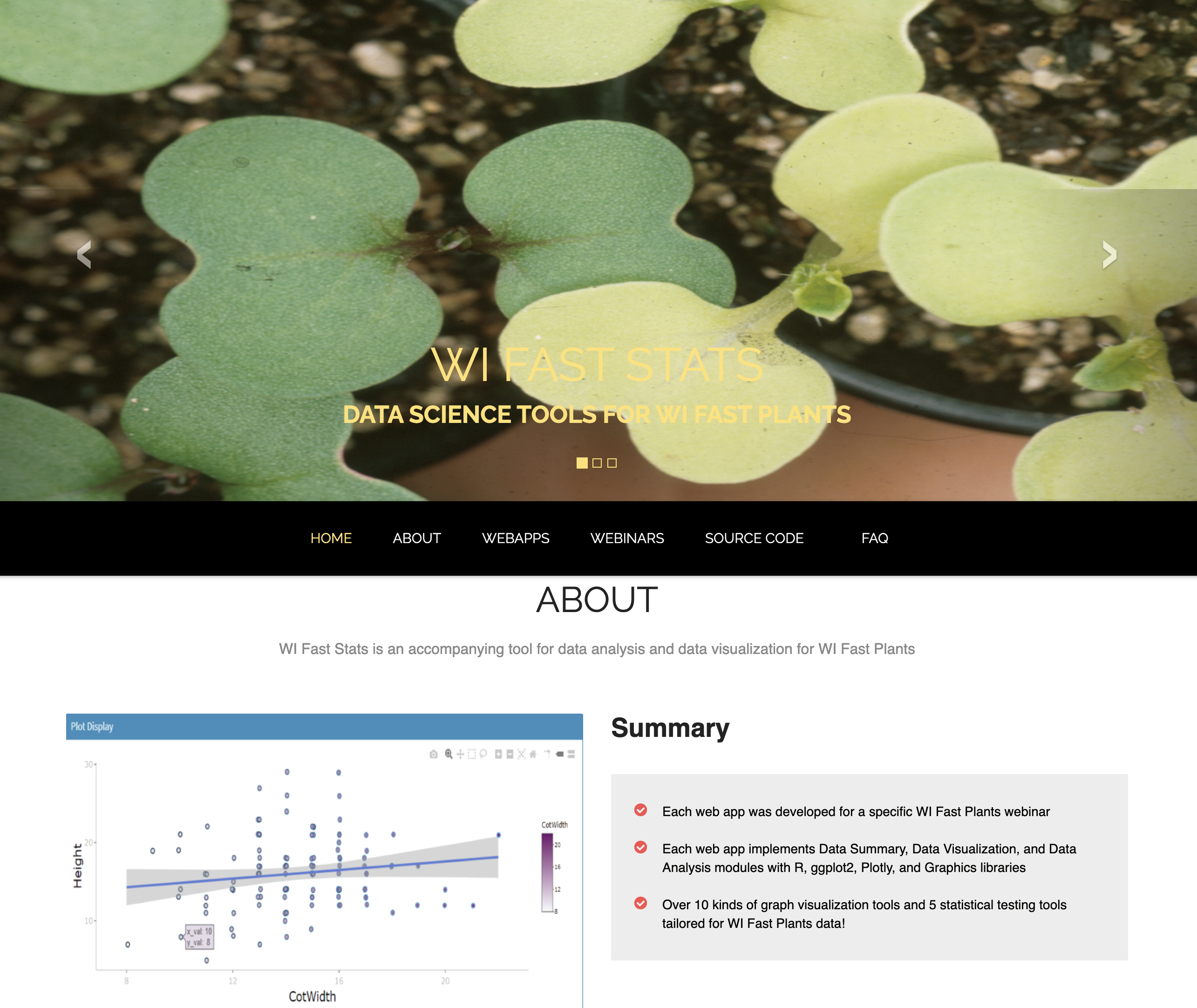}
    \caption{WI Fast Stats website home page: \url{https://wi-fast-stats.wid.wisc.edu/}.}
    \label{figweb1}
\end{figure}

\begin{figure}
    \centering
    \includegraphics[scale=0.25]{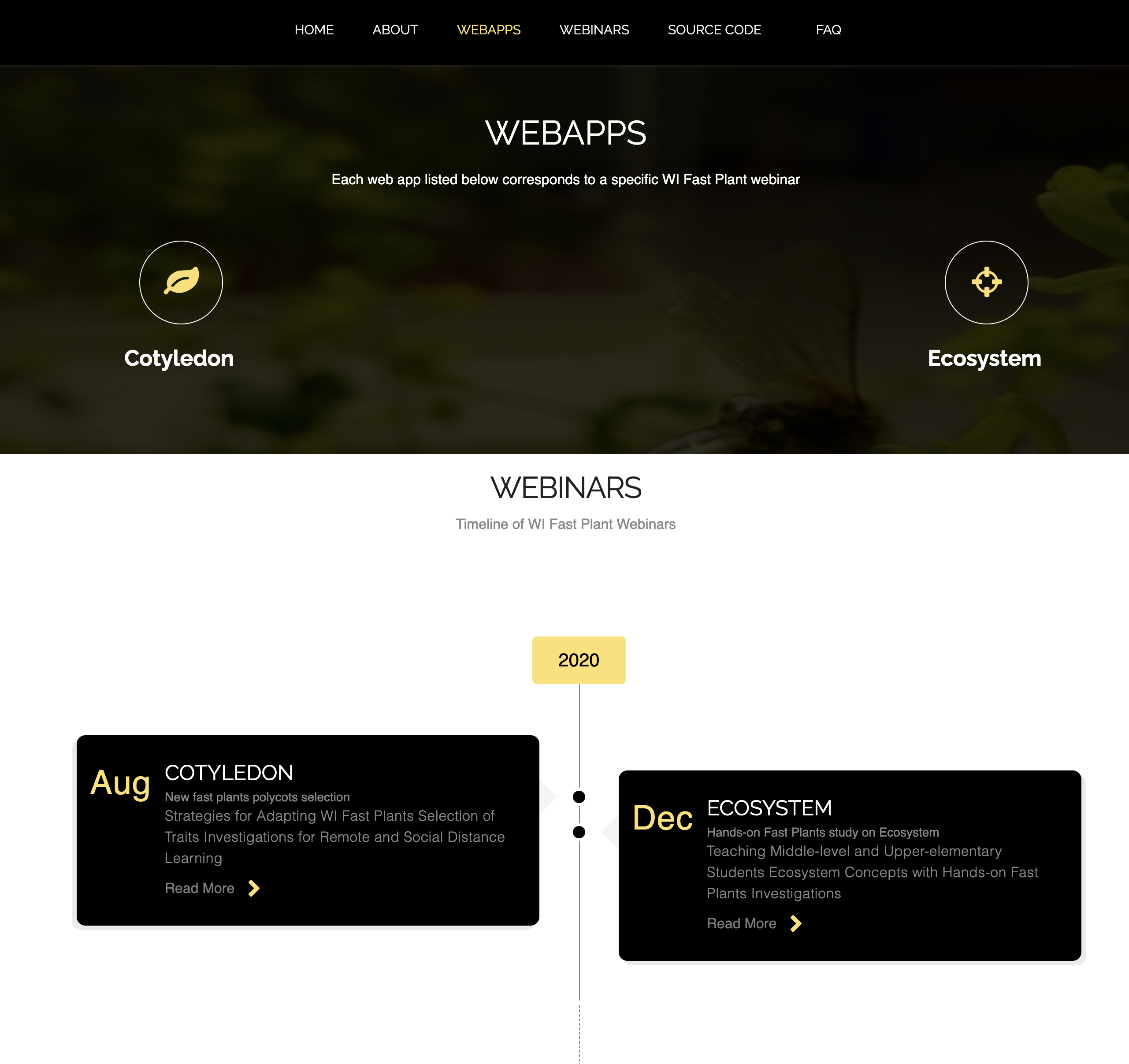}
    \caption{WI Fast Stats website comprises different web apps each corresponding to a given WI Fast Plants webinar.}
    \label{figweb2}
\end{figure}

WI Fast Stats (\url{https://wi-fast-stats.wid.wisc.edu/}) is an integrated animated web page which hosts a collection of web applications and relevant webinars for WI Fast Plants. The web page contains six modules including Home, About, Webapps, Webinars, Source Code, and FAQ (Figures \ref{figweb1} and \ref{figweb2}). The Home page utilizes \texttt{animate.css} to construct the animation with three scrolling images retrieved from WI Fast Plants. \texttt{Font-awesome.css} offers a fantastic mode to show buttons representing the links of our web applications as well as the Github source code. We also built a timeline to store all the webinar-related information.

The web page and accompanying web apps are all open-source with the code stored in the GitHub repository: \url{https://github.com/crsl4/fast-stats}.

\section{Description of the web apps}

\begin{figure}
    \centering
    \includegraphics[scale=0.25]{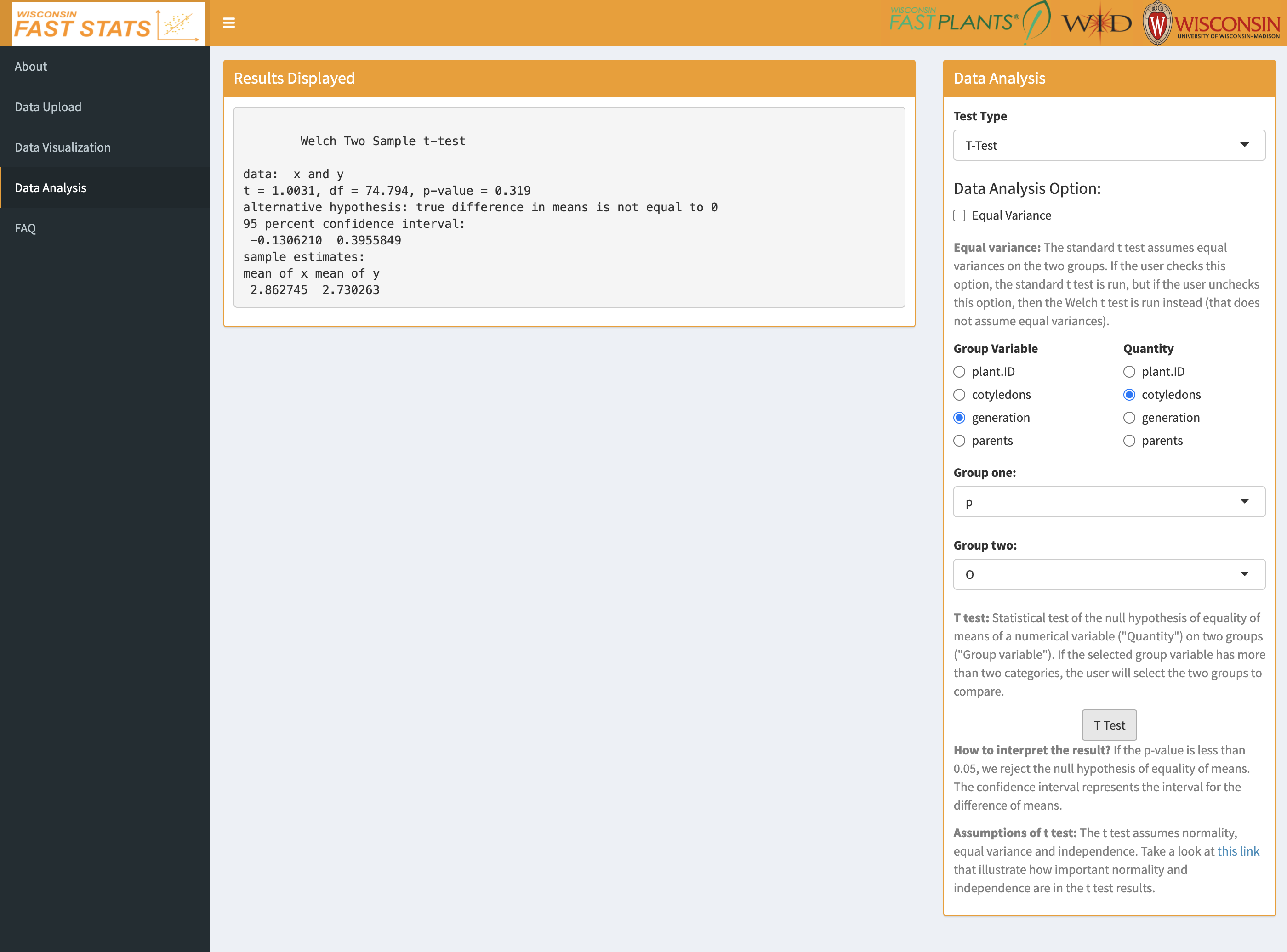}
    \caption{WI Fast Stats web app corresponding to the WI Fast Plants webinar on the selection of polycot plants: \url{https://wi-fast-stats.wid.wisc.edu/cotyledon/}.}
    \label{figwebapp1}
\end{figure}

\begin{figure}
    \centering
    \includegraphics[scale=0.25]{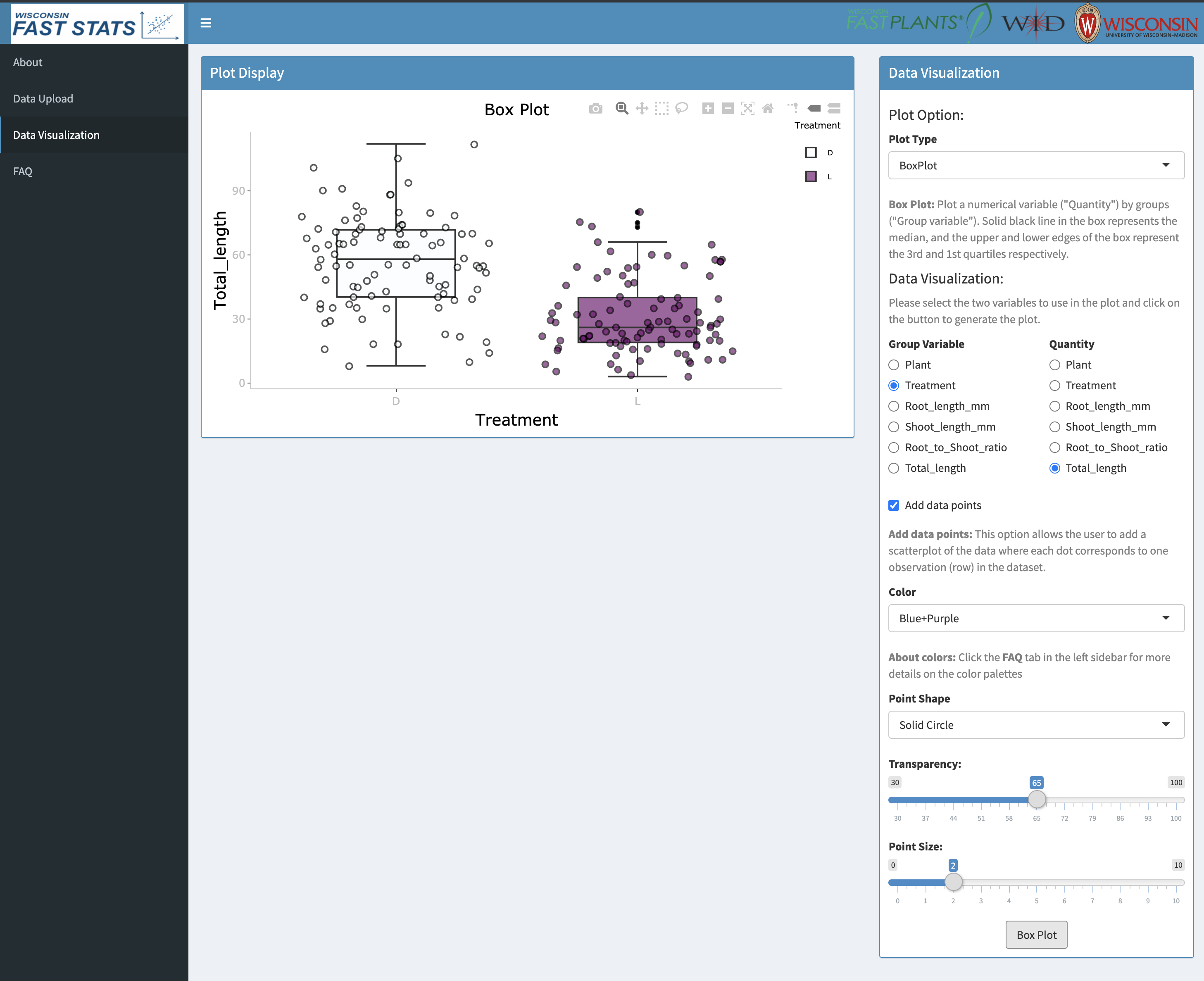}
    \caption{WI Fast Stats web app corresponding to the WI Fast Plants webinar on the effect of the ecosystem on the plants.: \url{https://wi-fast-stats.wid.wisc.edu/ecosystem/}.}
    \label{figwebapp2}
\end{figure}

The web apps within WI Fast Stats provide an interactive and easy-to-use platform for data visualization and data analysis for data collected in accordance to the WI Fast Plants webinars and educational materials.
The web apps are based on the \texttt{R shiny} package and contain three main modules: Data Summary, Data Visualization, and Data Analysis which are built with the \texttt{ShinyDashboard} framework (Figures \ref{figwebapp1} and \ref{figwebapp2}). 

The Data Upload section allows the users to upload their own collected data based on their fast plants experiments, or to utilize the already loaded sample dataset which illustrates the same educational outcomes intended in the webinar without having to run the experiments.
In addition, this section also provides a summary button to show only first rows of the dataset or the whole dataset.

The Data Visualization section allows the users to create five plots: Mosaic plot, Scatter plot, Box plot, Violin plot, and Density plot. It utilizes the \texttt{ggplot2} library to declaratively create graphs based on particular group variables and quantity variables. Furthermore, the \texttt{Plotly} library extends the features of the plot by adding \texttt{Lasso Select}, \texttt{autoscale}, and \texttt{data-toggle} tooltip. Finally, color palettes, transparency, point size, and point shape are available options to improve the overall appearance of the graph. 

The Data Analysis section (not present in web apps tailored to middle
school students) allows the user to perform statistical tests like the
chi-square test and the t test to compare the characteristics of the plants under different environmental or experimental settings.

Finally, the web apps maintain a validator system detecting any illegal actions done by users and providing meaningful error messages. The website and web apps are accompanied by a specialized Google user group (\texttt{wi-fast-stats}) for general questions.

\section{Future work}

WI Fast Stats will be continuously evolving to provide data visualization and data analysis capabilities for the ever-growing needs of the WI Fast Plants community.

\section*{Acknowledgements}
This work was supported by the Department of Energy [DE-SC0021016 to CSL]. We thank Hedi Baxter Lauffer and everybody at WI Fast Plants for inviting us to work with them on the creation of these Data Science educational open-source tools. Finally, we acknowledge the work in \cite{Hotaling2020} which helped us improve the scientific writing of this manuscript.

\bibliographystyle{unsrt}  

\bibliography{paper}

\end{document}